\documentclass[aps,epsfig,float,twocolumn]{revtex4}

\usepackage{graphicx}
\def\dsum#1#2{\displaystyle\sum_{#1}^{#2}}
\def\dfrac#1#2{\displaystyle\frac{#1}{#2}}
\def\<{\langle}
\def\>{\rangle}
\def\({\left(}
\def\){\right)}
\def\[{\left[}
\def\]{\right]}
\def\up{\uparrow}
\def\dn{\downarrow}

\def\e{\mathrm{e}}
\def\i{\mathrm{i}}
\def \IC {\mathrm{IC}}
\def \C {\mathrm{C}}
\def\Re#1{\mathrm{Re}\left\{#1\right\}}

\begin{document}
\title{Commensurate-incommensurate transition and associated collective modes in the stripe state of cuprates near $\frac{1}{8}$ hole-doping}
\author{E. Kaneshita\footnote{Present address:  Advanced Photon Source, Argonne National Laboratory, Argonne, IL 60439.}, I. Martin, and A. R. Bishop}
\affiliation{
Los Alamos National Laboratory, Los Alamos, NM 87545}
\date{\today}
\begin{abstract}

We analyze the anomalous behavior in hole-doped cuprates near $\frac{1}{8}$ doping in terms of the commensurate-incommensurate transition of a stripe phase.
Based on an effective Ginzburg-Landau theory appropriate for weak pinning, we calculate the commensurate-incommensurate transition point and the energies of the phason and amplitudon collective modes.
Using experimentally available parameters, we estimate the phason gap (pinning frequency), the conductivity and the contribution of the phason mode to the dielectric function.
\end{abstract}
\maketitle

\section{Introduction}
Many of the high-$T_c$ cuprate materials show signatures of static or dynamic spin and charge modulation, or \textit{stripes}~\cite{suzuki,tranquada1,tranquada2}.
The effect of stripes on high-temperature superconductivity is a subject of ongoing debate:
Some researchers consider them central to the mechanism of superconductivity, others consider them detrimental, and yet others argue that stripes are irrelevant.
The situation is further complicated by the difficulty of detecting the presence of stripes, particularly when they are only weakly pinned to the lattice, or dynamical.
The former case is typically associated with zero-frequency incommensurate-momentum response in neutron scattering~\cite{tranquada2}, while the latter is associated with finite-frequency incommensurate-momentum response~\cite{mook}.
The difficulty of direct detection has motivated other approaches to stripe detection based on their effects, e.g., on the phonon modes~\cite{mcqueeney}, or lattice structure~\cite{bozin,bianconi}.
In particular, one naturally expects that in the presence of sharp stripes there will be a rearrangement of the collective mode spectrum, with the formation of local modes split off from the bulk continuum, or \textit{edge modes}~\cite{yu,ivar}.

One of the most striking effects in the striped cuprates is the suppression of the superconductivity near the $\frac{1}{8}$ hole-doping, where (based on neutron scattering) the stripes become stationary: the so-called ``$\frac{1}{8}$-anomaly"~\cite{moodenbaugh}.
At $\frac{1}{8}$ doping, stripes are locked in with the lattice resulting in a commensurate modulation wavenumber.
This suppresses the charge and spin fluctuations, which are considered to be important for superconductivity.
Away from $\frac{1}{8}$ doping, on the other hand, the favorable wavenumber for the stripe structure is expected to be incommensurate and proportional to the hole concentration.
Naturally, the incommensurability renders the stripes more dynamical.
While this anomaly strongly suggests that the commensurate (C) stripes suppress superconductivity, the question of whether incommensurate (IC) (or fluctuating) stripes promote the superconductivity remains.
We believe that study of this interplay is important for developing an understanding of the high-temperature superconductivity in cuprates.
As a first step, in this paper, we discuss the C-IC transition near the commensurate doping, calculating the transition point within a weak pinning approximation.
We also calculate the energy of two charge-spin-lattice excitations, namely, the \textit{phason} and the \textit{amplitudon}.
In the last part of the paper, we provide relations between Ginzburg-Landau parameters (order parameters, energy coefficients) and physical quantities (conductivity, dielectric function) by considering the response to an electric field.

\section{Model}
We model the stripe phase by a simultaneous unidirectional sinusoidal charge density wave (CDW) and spin density wave (SDW) with wavelengths $\lambda=\frac{G}{q}a$ and $2\lambda$, respectively.
Here $a$ is the lattice constant, $G$ is the reciprocal-lattice vector, and $q$ is the ordering vector.
We introduce $\psi(x_i)$ and $\chi(x_i)$ as the complex order parameters for the CDW and the SDW, respectively.
The real part of $\psi(x)$ gives the deviation of the charge density from the uniform state:
\begin{eqnarray}
\psi(x)+\psi^*(x) = \rho(x) - \rho_0,
\end{eqnarray}
where $\rho_0$ is the average hole-density per site:
$\rho_0=z$ for a doping $z$.
The real part of $\chi(x)$ gives the staggered spin density:
\begin{eqnarray}
\chi(x)+\chi^*(x) = (-1)^{x/a}\sigma(x).
\end{eqnarray}

The periodicity of the CDW, $\nu$, is characterized by
\begin{eqnarray}
\nu \equiv \frac{\lambda}{a} = \frac{G}{q}.
\label{eq:commensurability}
\end{eqnarray}
If the CDW and lattice are commensurate, we can write $\nu=L/\tilde{L}$ with relatively prime integers $L$ and $\tilde{L}$.
We extend $\nu=L/\tilde{L}$ to irrational numbers for the IC case by taking $L \rightarrow \infty$ and $\tilde{L} \rightarrow \infty$.
We define $\frac{1}{L}$ as a commensurability; thus, the commensurability for the IC case is zero.
The order parameters have the following forms:
\begin{eqnarray}
&&\mbox{$\psi(x,\phi) = \rho \exp[\i(q x + \frac{\phi}{L})]$},
\label{eq:order-ch}\\
&&\mbox{$\chi(x,\phi) =  - \i m \exp[\i(\frac{q}{2} x + \frac{\phi}{2L})]$}.
\label{eq:order-spn}
\end{eqnarray}

We investigate the free energy near the commensurate doping $z_c=\frac{1}{8}$.
Here we consider only the following two cases as the groundstate candidates. One is the IC case,
where the periodicity is given by $\nu=\frac{1}{2z}$ ($L=\infty$) as a function
of the doping $z$. The other is the simple C case with $\nu=L$ (finite $L$)
independent of the doping $z$.
We denote $q$ especially in the IC case by $Q(z) \equiv 2zG$.

\begin{figure}[htbp]
\begin{center}
\includegraphics[width = 0.7\linewidth]{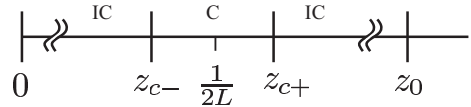}
\caption{The C-IC transition near the commensurate doping $z_c=\frac{1}{2L}$.} \label{fig:transition}
\end{center}
\end{figure}

We introduce the Ginzburg-Landau functional (per site):
\begin{eqnarray}
F= \int \frac{dx}{V_x}\, f[\psi(x),\chi(x)].
\end{eqnarray}
In general, $f$ can be written in the following form for a one-dimensional CDW with commensurability $\frac{1}{L}$~\cite{mcmillan1,mcmillan2}:
\begin{eqnarray}
f[\psi(x),\chi(x)] &=& f_0[\psi(x),\chi(x)]+ p(x) \Re{\psi(x)^L}  \nonumber\\
&&+ g_0\left|[Q(z)+\i\nabla]\psi(x)\right|^2.
\label{eq:f}
\end{eqnarray}
Here, $f_0$ is concerned with the spin and charge ordering, but not with the pinning or deformation of the CDW.
The second term in expression (\ref{eq:f}) is introduced to represent the pinning energy of the CDW.
The third term, which originates from the gradient term in the free energy, gives the preference for $Q(z)$ ($g_0>0$).
We introduce this term to describe the energy which arises from the deformation of the CDW;
however, to simplify the calculation, it does not include the local effects (e.g., those from discommensurations~\cite{mcmillan2}) but only gives the preference among different wavevectors.
We assume that $f_0$ is of the form
\begin{eqnarray}
f_0[\psi(x),\chi(x)] &=& r_0 \left|\chi(x)\right|^2 + u_0 \left|\chi(x)\right|^4\nonumber\\
&&-s_0\,\left|\chi(x)^2\psi(x)\right|+v_0 \left|\psi(x)\right|^2,
\end{eqnarray}
with the parameters chosen in such a way that the charge order is induced by the magnetic order~\cite{pryadko}.
Expanding $p(x)$ in harmonics,
\begin{eqnarray}
p(x)= \sum_l p_l \cos(lGx),
\end{eqnarray}
the free energy per site can be rewritten as
\begin{eqnarray}
F(\rho,m,\phi)
&=&F^{(0)}(\rho,m)
+\tilde{p}_1\,\rho^L \nonumber\\
&&+ \tilde{g}_0[Q(z)-q]^2 \rho^2,
\label{eq:F_phi}\\
F^{(0)}(\rho,m)&=&
r_0 m^2+u_0m^4-s_0m^2\rho
+v_0\rho^2,
\label{eq:f0}
\end{eqnarray}
where
\begin{eqnarray}
\tilde{p}_1=\frac{1}{2}p_1\cos(\phi_0),\hspace{0.3cm}
\tilde{g}_0=\frac{1}{2}g_0.
\end{eqnarray}
For the equilibrium state, $\phi_0$ is taken so that $p_1\cos(\phi_0)=-|p_1|$
(i.e.,  $\tilde{p}_1$ is always negative, and the sign of $p_1$
determines the location of the stripe center). Notice that the pinning term disappears for incommensurate cases, since $\rho^L\rightarrow 0$ for
$L\rightarrow \infty$.
The $r_0$ and $u_0$ terms govern the continuous transition between the stripe and the uniform states at a large doping. Since
we are interested in the stripe state, we take $r_0<0$, $u_0>0$, $s_0>0$ and
$v_0>0$. $r_0(z)$ vanishes at the order-disorder transition point $z_0$. The
region between the C-IC transition points $z_{c-}<z<z_{c+}$ around the C point $\frac{1}{2L}$ is the C region; outside this region the system is in the IC
phase (Fig.~\ref{fig:transition}).
We will find in the following that the width of the C region rapidly shrinks with increasing $L$.

Although the term $\rho^L$ can give large negative energy for large $\rho$,
this does not cause any problem so long as we restrict $\rho_\C$ to not too
large values, so that the $\rho_{\C}^L$ term is not dominant. In other words,
this term makes only a minor modification of the free energy, which selects
between the C and IC equilibrium states.

We make the following \textit{assumptions} in this paper:
\begin{enumerate}
\item $|\tilde{p}_1| \rho^{L-2} \ll v_0$ (weak pinning).
\item $s_0^2 \ll u_0 v_0$ (weak coupling).
\item $r_0$ is constant around the C region ($z_{c\pm} \ll z_0$).
\item $m < 0.1$.
\item $\Lambda (\equiv\frac{\rho}{m^2}) \lesssim 1$
(equivalent to $s_0 \lesssim v_0$).
\end{enumerate}
\textit{Assumption 4} is consistent with the neutron scattering experiments
where it is roughly estimated that $2m\approx0.1$ for striped
cuprates~\cite{kimura}.
In \textit{assumption 5}, the inequality sign stands for ``to be on the order of or less than", and the \textit{assumption} reflects the fact that the charge order is much weaker than the magnetic one in metallic stripes due to the strong fluctuations.

\section{Relative stability of C and IC phases}
Now we evaluate the order parameters for the C and IC cases so that $\rho$ and $m$ minimize the free energy.
First, in the IC case, it follows from $\frac{\partial F}{\partial m}=\frac{\partial F}{\partial \rho}=0$ and
$F_{\IC}=F^{(0)}$ that
\begin{eqnarray}
m_{\IC}^2&=&\frac{|r_0|+s_0\rho_{\IC}}{2u_0}\approx\frac{|r_0|}{2u_0},
\label{eq:mIC}\\
\rho_{\IC}&=&\frac{|r_0|s_0}{4u_0v_0-s_0^2} = \frac{m_{\IC}^2s_0}{2v_0}.
\label{eq:rhoIC}
\end{eqnarray}
The following condition is required for $m_{\IC}^2>0$ and $\rho_{\IC}>0$:
\begin{eqnarray}
s_0^2 < 4 u_0 v_0,
\end{eqnarray}
which is satisfied by \textit{assumption} \textit{2}.
Substituting Eqs.~(\ref{eq:mIC}) and (\ref{eq:rhoIC}) into Eq.~(\ref{eq:f0}), we obtain the condensation energy
\begin{eqnarray}
E_{con}=-\frac{1}{2}|r_0|m_\IC^2.
\end{eqnarray}

Next, in the C case, the magnetic order parameter is
\begin{eqnarray}
m_{\C}^2=\frac{|r_0|+s_0\rho_{\C}}{2u_0}\approx m_{\IC}^2.
\label{eq:mC}
\end{eqnarray}
The charge order parameter for the C case satisfies the equation
\begin{eqnarray}
\rho_{\C}^{L-1} + A\rho_{\C}+ B=0,
\label{eq:rho_c}
\end{eqnarray}
where
\begin{eqnarray}
A&=&-\dfrac{2v_0 - \frac{s_0^2}{2u_0}
 + 8\tilde{g}_0G^2\(z-\frac{q}{2G}\)^2 }{L\,|\tilde{p}_1|},\\
B&=&\dfrac{|r_0|\,s_0}{2u_0L\,|\tilde{p}_1|}.
\end{eqnarray}
Since $B>0$ and $A<0$, the following condition is required for the existence of a positive root:
\begin{eqnarray}
B-(L-2)\(\frac{|A|}{L-1}\)^{\frac{L-1}{L-2}}<0.
\label{eq:condition}
\end{eqnarray}
Under this condition, Eq.~(\ref{eq:rho_c}) has two positive roots.
The smaller one corresponds to the minimal energy in the C case, while the other corresponds to the maximal one; therefore $\rho_{\C}$ is given by the smaller positive root of Eq.~(\ref{eq:rho_c}).

$A$ is a function of $z$ and $L$. For later use, we define $A_0$ and $B_0$
independent of $z$ and $L$ near the commensurate point:
\begin{eqnarray}
A_0 &\equiv& L\,|\tilde{p}_1|\,A + 2Z^2
=-\frac{4u_0v_0 - s_0^2}{2u_0},\\
B_0 &\equiv& L\,|\tilde{p}_1|\,B
=\dfrac{|r_0|s_0}{2u_0},
\end{eqnarray}
where
\begin{eqnarray}
Z^2 &\equiv& 4\tilde{g}_0G^2\(z-\frac{q}{2G}\)^2.
\label{eq:Z}
\end{eqnarray}
Note that, near $z_0$, $B_0$ depends on $z$ through $r_0$, but not so far below $z_0$ (from \textit{assumption 3}).
From now on, we also use the notation $Z$ with a subscript to represent that corresponding to $z$ with the same subscript through Eq.~(\ref{eq:Z}).

We discuss the C-IC transition as a function of the hole concentration $z$.
We will evaluate the difference of the free energy between C and IC below.
However, $F_{\IC}$ is not well-defined at $Z=0$ since there is no IC state at $Z=0$.
To avoid problems with this discontinuity, we first define
\begin{eqnarray}
\rho_{\IC}&\equiv&\frac{B_0}{|A_0|}=\frac{|r_0|s_0}{4u_0v_0-s_0^2},\\
F_{\IC}&=&F^{(0)}(\rho_{\IC})
\end{eqnarray}
as the order parameter and the IC free energy for any $Z$ including $Z=0$, and then we take the commensurability at the $Z=0$ point into account after comparing the continuous free-energy functions if needed.

The difference between the C and IC free energies is [from Eqs. (\ref{eq:mIC}), (\ref{eq:mC}), and (\ref{eq:rho_c})]
\begin{eqnarray}
\Delta F &\equiv& F_{\mathrm{\IC}}-F_{\mathrm{C}}
=\Delta F^{(0)}
+|\tilde{p}_1| \rho_{\C}^L - Z^2 \rho_{\C}^2\\
&=&
-\frac{|A_0|}{2}\(1-\frac{2}{L}\)\(1+\tilde{Z}^2\) \rho_{\C}^2\nonumber\\
&&{}+\(1-\frac{1}{L}\)\,|A_0|\,\rho_{\IC}\rho_{\C}
-\frac{|A_0|}{2}\rho_{\IC}^2,
\label{eq:DF2}
\end{eqnarray}
where $\tilde{Z^2}=\frac{2Z^2}{|A_0|}$.

The roots of $\Delta F=0$ for $L\ge3$ are
\begin{eqnarray}
\rho_{\C}^{(\Delta F=0)}
=\frac{L-1\pm\sqrt{1-L(L-2)\tilde{Z}^2}}{(L-2)(1+\tilde{Z}^2)}\rho_{\IC}.
\end{eqnarray}
Here the smaller root gives the C-IC transition, because $\frac{d (\Delta F) }{d\rho_\C}$ should be positive so that
$\frac{d (\Delta F) }{d(\tilde{Z}^2)} < 0$
at the transition point.
We require $\tilde{Z}_c^2<\frac{1}{L(L-2)}$ so that the C-IC transition exists.
For $L\ge3$, $\rho_{\C}^{(\Delta F=0)}$ has only one minimum at $\tilde{Z}=0$.
We find that $\rho_{\C}$ is larger than $\rho_{\IC}$ at $Z_c$ ($<Z_0$):
\begin{eqnarray}
\rho_{\C}^{(\Delta F=0)}
=\rho_{\IC}\[1+\frac{L-2}{2}\tilde{Z}^2
+\cdots\]>\rho_{\IC}.
\label{eq:rho_c(dF=0)}
\end{eqnarray}

From Eq.~(\ref{eq:rho_c}), we can write
\begin{eqnarray}
\rho_{\C}&=&  \frac{\rho_{\IC}}{(1+\tilde{Z}^2)} +\frac{\rho_{\C}^{L-1}}{|A|}.
\label{eq:rho_c2}
\end{eqnarray}
$\rho_{\C}$ and $\rho_{\IC}$ vanish together at the order-disorder transition point $Z=Z_0$.
Note that we take $\rho_{\IC}$ as a constant for $Z$ far from $Z_0$, but it is a function of $Z$ near $Z_0$ which vanishes at $Z=Z_0$.
The order parameter at the C-IC phase transition, $\rho_{\IC}(Z_c)$ ($\neq0$), should satisfy both Eqs.~(\ref{eq:rho_c(dF=0)}) and (\ref{eq:rho_c2}).

For $\tilde{Z}_c^4 \ll 1$, from Eq.~(\ref{eq:rho_c(dF=0)}),
\begin{eqnarray}
\(\rho_{\C}^{(\Delta F=0)}\)^{L-1}
&\approx&\rho_{\IC}^{L-1}\[1+\frac{(L-2)(L-1)}{2}\tilde{Z}^2\].
\nonumber\\
\end{eqnarray}
Substituting this and Eq.~({\ref{eq:rho_c(dF=0)}}) into Eq.~({\ref{eq:rho_c2}}) and keeping terms up to the order in $\tilde{Z}_c^2$,
\begin{eqnarray}
1+\frac{L-2}{2}\tilde{Z}_c^2 &\approx&\(1+LX\)
+\[\frac{L^2(L-3)}{2}X-1\]\tilde{Z}_c^2\nonumber\\
\tilde{Z}_c^2
&\approx&\frac{2X}{1-L(L-3)X},
\end{eqnarray}
where $X\equiv\rho_{\IC}^{L-2}\frac{|\tilde{p}_1|}{|A_0|}$.

Since $X^2\sim0$ from \textit{assumptions 1} and \textit{2}, $\tilde{Z}_c^2$ ($=\frac{2Z_c^2}{|A_0|}$) is represented by a simple function of $|\tilde{p}_1|$:
\begin{eqnarray}
\tilde{Z}_c^2 \approx 2\rho_{\IC}^{L-2}\frac{|\tilde{p}_1|}{|A_0|}.
\label{eq:transition}
\end{eqnarray}
By virtue of \textit{assumption} \textit{1}, it is apparent that $\tilde{Z}_c^4 \ll 1$, which is consistent with the case we consider here; therefore it is confirmed that $\tilde{Z}_c$ in Eq.~(\ref{eq:transition}) is the C-IC transition point for the case where \textit{assumptions 1} and \textit{2} hold.

The width of the commensurate region is given by
\begin{eqnarray}
z_{c+}-z_{c-} \approx \frac{1}{G} \sqrt{\frac{|\tilde{p}_1|}{\tilde{g}_0}} \rho_{\IC}^{\frac{L-2}{2}}.
\end{eqnarray}
The commensurate region becomes exponentially narrow as the CDW period $L$ increases.
Therefore, only commensurate stripes with the relatively small L may exist in practice.
Note that, here, we do not consider the case where discommensurations exist~\cite{mcmillan2}.

At a commensurate point, the condensation energy is given by
\begin{eqnarray}
E_{con}&=&-E_{con}^{(0)}-E_{\C},\\
E_{con}^{(0)}&=&\frac{1}{2}|r_0|m_{\C}^2,\\
E_{\C}&=&\frac{L-1}{2}|\tilde{p}_1|\rho^L.
\end{eqnarray}

\section{Collective modes: phasons and amplitudons}
Now we consider the slightly displaced CDW to study collective excitations around the equilibrium state.
First, we discuss the \textit{phason} and then the \textit{amplitudon}.

For the \textit{phason}, we introduce an $x$-dependence into the phase of the order parameters:
\begin{eqnarray}
&&\mbox{$\psi(x,\phi) = \rho \exp\[\i(q x +\frac{\phi_0}{L}+ \frac{\varphi(x)}{L})\]$},
\label{eq:psi-phason}\\
&&\mbox{$\chi(x,\phi) =  -\i m \exp\[\i(\frac{q}{2} x +\frac{\phi_0}{2L}+ \frac{\varphi(x)}{L})\]$}\\
&& \varphi(x)= \xi \cos(kx+\phi_1),
\label{eq:phase}
\end{eqnarray}
where $k=\frac{q}{N}$ with $N\ge2$.
The free energy is given by
\begin{widetext}
\begin{eqnarray}
F(\rho,\xi)&=&
\int_V\frac{dx}{V}\left\{
f_0[\psi(x),\chi(x)]+ p(x)\Re{\psi(x)^L}
+ g_0\left|Q(z)-q+\frac{k\xi\sin(kx+\phi_1)}{L}\right|^2|\psi(x)|^2\right\}\\
& \approx &F_0(\rho)+F^{(\varphi)}_2(\rho)\xi^2,\\
F_0(\rho)&=& F^{(0)}(\rho)
-|\tilde{p}_1|\,\rho^L
+ \tilde{g}_0[Q(z)-q]^2 \rho^2,\\
F^{(\varphi)}_2(\rho)&=&
\frac{1}{2}\,|\tilde{p}_1|\,\rho^L\frac{1+\delta_{k,0}\cos(2\phi_1)}{2}
+ \frac{1}{2}\frac{\tilde{g}_0\rho^2}{L^2}k^2
=\frac{1+\delta_{k,0}\cos(2\phi_1)}{2}\frac{\rho^2}{L^2}
\[\frac{L^2}{2}\,|\tilde{p}_1|\,\rho^{L-2}
+ \tilde{g}_0 k^2\],
\end{eqnarray}
\end{widetext}
Assuming that $\rho$ depends on time $t$ only through $\xi(t)$, the kinetic energy is given by
\begin{eqnarray}
T &=& t_0 \int \frac{dx}{V}\left|\frac{d\psi}{dt}\right|^2\\
&=&\frac{t_0 \rho^2}{L^2} \frac{1+\delta_{k,0}\cos(2\phi_1)}{2}\(\frac{d\xi}{dt}\)^2,
\end{eqnarray}
where $t_0$ is a Ginzburg-Landau parameter of the kinetic energy and defined by this equation.
Thus, the Lagrangian corresponding to the fluctuating CDW is
\begin{eqnarray}
\mathcal{L} &=& \mbox{$\frac{1}{2}\[1+\delta_{k,0}\cos(2\phi_1)\] \frac{t_0 \rho^2}{L^2}$}\nonumber\\&&\mbox{$\times\[
 \(\frac{d\xi}{dt}\)^2
-\frac{1}{t_0} \(
\frac{L^2}{2}\,|\tilde{p}_1|\,\rho^{L-2}
+ \tilde{g}_0k^2\)\xi^2 \]$}\nonumber\\
&&{}+\mathrm{const.}
\end{eqnarray}
Solving the Euler-Lagrange equation of motion,
\begin{eqnarray}
\xi(t)&\propto&\cos(\Omega_{\varphi} t +\phi_2),\\
\Omega_{\varphi}(k)^2&=&
\frac{1}{t_0}\[\frac{L^2}{2}
|\tilde{p}_1|\rho^{L-2}
+ \tilde{g}_0 k^2\].
\end{eqnarray}

\begin{figure}[htbp]
\begin{center}
\includegraphics[width = 0.8\linewidth]{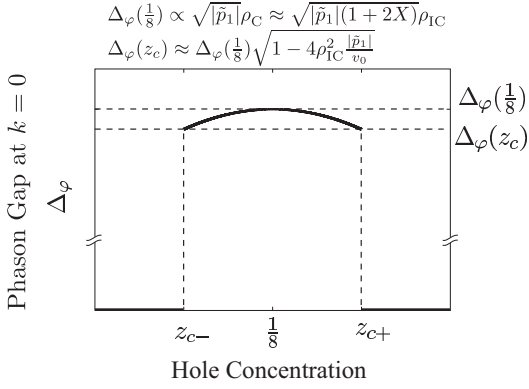}
\caption{The phason gap around $\frac{1}{8}$ doping ($L=4$).
The width of the commensurate region is $z_{c+}-z_{c-}\approx\frac{a}{4\pi} \sqrt{\frac{|\tilde{p}_1|}{\tilde{g}_0}} \frac{|r_0|\,s_0}{u_0v_0}$.
}\label{fig:gap}
\end{center}
\end{figure}

$\Omega_\varphi$ is gapless in the IC case and gapped in the C case.
From Eq.~(\ref{eq:rho_c2}), the gap at $k=0$ can be evaluated as
\begin{eqnarray}
\Delta_\varphi^2
&\equiv&\frac{L^2}{2t_0}|\tilde{p}_1|\rho_{\C}^{L-2}
\\
&\approx& \frac{L^2}{2t_0}|\tilde{p}_1|
\rho_{\IC}^{L-2}\big[1+(L-2)X\big]
\big[1-(L-2)\tilde{Z}^2\big],\nonumber\\
\end{eqnarray}
where we have used $\tilde{Z}^4<\tilde{Z}_c^4\ll1$ in the commensurate region [see Eq.~(\ref{eq:transition})].
In Fig.~\ref{fig:gap}, the doping dependence of the phason gap is shown for the $L=4$ case.

In Appendix \ref{app:GL-estimate}, $t_0$ is roughly estimated as $\sim \rho_\C^{-2} \times 10^{-32}$ $\mbox{[eV s$^2$]}$ using a single-band Peierls-Hubbard model.
\begin{eqnarray}
\Delta_\varphi(\mbox{$\frac{1}{8}$})
&\sim& \sqrt{E_{\C}} \times 10^{16} \,\mbox{[Hz]},
\end{eqnarray}
where $E_\C$ ($=|\tilde{p}_1|\rho^4$) is the commensurability energy.
$E_{\C}$ is roughly estimated as $U^7m^8 \times 10^{-2}$ [eV] in Appendix~\ref{app:GL-estimate} ($U$ is the on-site Coulomb interaction in the single-band Peierls-Hubbard model).
Therefore, the order of the phason gap at $z=\frac{1}{8}$ is
\begin{eqnarray}
\Delta_\varphi(\mbox{$\frac{1}{8}$})
&\sim& U^\frac{7}{2}m^4\times10^{15}\,\mbox{[Hz]}
\sim U^\frac{7}{2}m^4\,\mbox{[eV]}.
\label{eq:phason-gap}
\end{eqnarray}
We expect that $U < 10$ [eV] and $m<0.1$ (\textit{assumption 4}), and thus find the order of $\Delta_\varphi(\frac{1}{8})$ is less than $100$ [meV].
For a smaller $U$, it can be even much smaller than $100$ [meV]:  e.g. for $U\approx1$ [eV], we find $\Delta_\varphi(\frac{1}{8})< 1$ [meV].

Next we consider the \textit{amplitudon} mode.
For this purpose, we write the order parameters in the following form:
\begin{eqnarray}
&&\mbox{$\psi(x,\phi) = \rho\[1+\mathcal{A}(x)\] \exp\[\i(q x + \frac{\phi_0}{L})\]$},\\
&&\mbox{$\chi(x,\phi) =  -\i m\[1+\mathcal{A}(x)\] \exp\[\i(\frac{q}{2} x + \frac{\phi_0}{2L})\]$}\\
&& \mathcal{A}(x)=\zeta \cos(kx+\phi_1).
\end{eqnarray}
Then, the free energy is given by
\begin{widetext}
\begin{eqnarray}
F(\rho,\zeta)&=&
\int_V\frac{dx}{V}\left\{
f_0[\psi(x),\chi(x)]+ p(x)\Re{\psi(x)^L}
+ g_0\[Q(z)-q\]^2|\psi(x)|^2
+ g_0\[k\zeta\sin(kx)\]^2\rho^2\right\}\\
&\approx&F_0(\rho)+F^{(\mathcal{A})}_2(\rho)\,\zeta^2,\\
F^{(\mathcal{A})}_2(\rho)&=&
\frac{1+\delta_{k,0}\cos(2\phi_1)}{2}
\Bigg\{
\frac{|r_0|^2}{u_0}
+\frac{|r_0|\,s_0}{u_0}\rho
+\Big(v_0+ \tilde{g}_0[Q(z)-q]^2\Big) \rho^2
+\frac{L(L-1)}{2}\,|\tilde{p}_1|\,\rho^{L}
+ \tilde{g}_0k^2\rho^2\Bigg\}.
\end{eqnarray}
\end{widetext}
Thus, we find the frequency of the \textit{amplitudon} fluctuation as
\begin{eqnarray}
\Omega_\mathcal{A}(k)^2 &=&
\frac{1}{t_0}\[
\frac{|r_0|^2}{u_0}\rho^{-2}
+\frac{|r_0|\,s_0}{u_0}
\rho^{-1}\right.\nonumber\\
&&\hspace{.5cm}{}+v_0
+ \tilde{g}_0[Q(z)-q]^2\nonumber\\
&&\left.\hspace{.5cm}{}+\frac{L(L-1)}{2}
|\tilde{p}_1|\rho^{L-2}
+ \tilde{g}_0k^2\]\\
&\equiv&\Delta_{\mathcal{A}}^2
+ \frac{\tilde{g}_0}{t_0}k^2
\end{eqnarray}
The \textit{amplitudon} has a gap even in the IC case (unlike the \textit{phason}).
The \textit{amplitudon} gap at $\frac{1}{8}$ doping is
\begin{eqnarray}
\Delta_{\mathcal{A}}
&\approx& \sqrt{\frac{2 |r_0| m^2 + v_0\rho^2}{t_0\rho^{2}}}\\
&\sim& \mbox{$\sqrt{ E_{con}^{(0)}}\times10^{16}$ [Hz]}\\
&\lesssim& \mbox{$U m\times10^{16}$ [Hz]},
\end{eqnarray}
where $U$ is in the unit of eV, and $U\ll10$ [eV] (see Appendix~\ref{app:GL-estimate}).

So far, we have discussed the one-dimensional collective modes, that is the
modes that are uniform in the direction along stripes. Next we consider the
\textit{meandering} mode of stripes, i.e. the transverse mode with the
wavevector oriented along the stripe direction.  This case is a simple
extension of that for the one-dimensional phason. The only difference is that
now we impose a $y$-dependence on $\xi$ and introduce the energy cost
associated with the spatial variation in the $y$-direction. Then, we obtain the
following Lagrangian
\begin{eqnarray}
\mathcal{L} &=&
\frac{1}{2}\[1+\delta_{k_x,0}\cos(2\phi_1)\] \frac{t_0 \rho^2}{L^2}\nonumber\\
&& \times\int \frac{dy}{V_y}\[
 \(\frac{d\xi}{dt}\)^2
-\frac{\tilde{h}_0}{t_0}\(\frac{d \xi}{dy}\)^2\right.\nonumber\\
&&\left.\hspace{1.5cm}{}-\frac{1}{t_0} \(
\frac{L^2}{2}\,|\tilde{p}_1|\,\rho^{L-2}
+ \tilde{g}_0k_x^2\)\xi^2 \]\nonumber\\
&&{}+\mathrm{const.},
\label{eq:L-meandering}
\end{eqnarray}
where $\tilde{h}_0$ is the energy coefficient of the $y$-derivative term.
Solving the Euler-Lagrange equation of motion, we find
\begin{eqnarray}
\xi(y,t) &\propto& \cos(k_y y+\Omega_m t),\\
\[\Omega_m(k_x, k_y)\]^2&=&
\frac{1}{t_0} \[\tilde{h}_0 k_y^2 + \tilde{g}_0k_x^2\]
+\Delta_\varphi^2.
\end{eqnarray}

In a similar manner, we find the two-dimensional amplitudon as
\begin{eqnarray}
\zeta(y,t)&\propto&\cos(k_y y+\Omega_\mathcal{A} t),\\
\[\Omega_\mathcal{A}(k_x,k_y)\]^2&=&
\frac{1}{t_0}\[\tilde{h}_0k_y^2
+ \tilde{g}_0 k_x^2\]
+\Delta_\mathcal{A}^2.
\end{eqnarray}

\section{Response to electric field: conductivity and dielectric function}
\label{sec:electric-field} To investigate collective oscillation of stripes
which can be experimentally examined, we consider the \textit{meandering} mode
contribution to the conductivity. Since we do not include impurity pinning
effects in this paper, our treatment is equivalent to the classical particle
model in the one-dimensional case. In the presence of an electric field
$E=E_0\exp[\i (ky+\omega t)]$ along the $x$ direction, the equation of motion
is given by (see Appendix~\ref{app:eq-motion} or reference~\cite{gruner})
\begin{eqnarray}
&&\frac{d^2\xi(y,t)}{dt^2}+\Gamma\frac{d\xi(y,t)}{dt}
-\frac{\tilde{h}_0}{t_0}\frac{d^2 \xi(y,t)}{dy^2}
+\Delta_\phi^2\xi(y,t)\nonumber\\
&&\hspace{4cm}=\frac{eG}{\mu}E_0\exp[\i (ky+\omega t)],
\end{eqnarray}
where $\mu$ is the effective mass of the CDW (see Appendix~\ref{app:GL-estimate}). The particular solution of this equation is
\begin{eqnarray}
\xi(y,t)&=&\frac{e \rho_0 G}{\mu}\frac{E}
{[\Omega_m(0,k)]^2-\omega^2+\i\omega\Gamma}.
\end{eqnarray}
Therefore, an induced current-density is
\begin{eqnarray}
j_m(y,t) &=& \frac{e (\rho_0/a^2c)}{G} \frac{d\xi}{dt}
=\frac{1}{ 4 \pi}
\frac{\i \omega \omega_p^2 E(y,t)}
{\Omega_m^2-\omega^2+\i\omega\Gamma},
\end{eqnarray}
where
\begin{eqnarray}
\omega_p^2=\frac{4 \pi \rho_0 e^2}{a^2 c \mu} \,\,\,\mbox{[C$^2$ m$^{-3}$ kg$^{-1}$]},
\end{eqnarray}
and $c$ is the lattice constant in the $z$-direction.

At $\frac{1}{8}$ doping, from the effective mass ($\mu \approx 9.2 \times 10^{-31}$ [kg]) estimated in Appendix~\ref{app:GL-estimate}, it follows
\begin{eqnarray}
\omega_p&\approx&\sqrt{ \frac{1.2}{9.2}} \times 10^{15} \,\,\mbox{[Hz]} \sim 10^{14}\,\,\mbox{[Hz]}.
\label{eq:plasma}
\end{eqnarray}
Once we know $\omega_p$ accurately, the charge order parameter can be
determined. The value of $\omega_p$ will be determined from the phason
contribution to the dielectric function and the phason gap, $\Delta_\varphi$
(see below).

The conductivity is given by
\begin{eqnarray}
\sigma(k,\omega)&\equiv&\frac{d\,j_m}{dE}
=\frac{1}{ 4 \pi }
\frac{\i \omega \omega_p^2}
{\Omega_m^2-\omega^2+\i\omega\Gamma}.
\end{eqnarray}
The contribution of the \textit{meandering} mode to the dielectric function is
\begin{eqnarray}
\epsilon_m(k,\omega)
=\frac{4\pi}{\i\omega}\sigma(k,\omega)
=\frac{\omega_p^2}{\Omega_m^2-\omega^2+\i\omega\Gamma}.
\end{eqnarray}
The contribution of the \textit{phason} mode to the static dielectric constant is
\begin{eqnarray}
\epsilon_\varphi(\omega=0)
=\frac{\omega_p^2}{\Delta_\varphi^2}=\frac{36\pi \rho_0^2 e^2\times10^9}{(aq)^2 c L^2 E_\C},
\end{eqnarray}
where $E_\C$ is in unit of hertz.
From the experimental data for $\epsilon_\varphi(\omega=0)$, $E_\C$ can be determined.
We can also determine $\omega_p$ from $\epsilon_\varphi(\omega=0)$ and $\Delta_\phi(\frac{1}{8})$, and the effective mass can be estimated from $\omega_p$ (this would be close to that of a free electron).

At $\frac{1}{8}$ doping, it follows from the estimated phason gap (\ref{eq:phason-gap}) and plasma frequency of the CDW (\ref{eq:plasma}) that
\begin{eqnarray}
\epsilon_\varphi(\omega=0) \sim U^{-7}\times m^{-8}\times10^{-1},
\end{eqnarray}
where $U$ is in unit of eV.

\section{Summary}
In this paper, we investigated the competition between C and IC stripe phases by calculating the C-IC transition point in the weak-pinning approximation.
This allowed us to estimate the strength of the lock-in effect, and evaluate the phason and amplitudon fluctuation frequencies, the ac conductivity, and the dielectric function.

Our calculation suggests a narrow C region around $\frac{1}{8}$ doping. In this region, there is a weakly-pinned \textit{phason} mode (with a
charge-order-dependent gap at $k=0$).
The phason gap at $k=0$ for $\frac{1}{8}$ doping is estimated to be of the order of $10$ meV.
The pinning frequency increases with $|k|$, quadratically near $|k|=0$ and linearly for large $|k|$.
This weakly-pinned behavior should be observable by electronic transport
experiments. In the IC regime (away from $\frac{1}{8}$ doping), the phason
frequency is linear in $|k|$ with no gap at $k=0$. The \textit{amplitudon}
shows a similar $k$-dependent frequency as the \textit{phason}. The
\textit{amplitudon} gap at $k=0$ in the C phase is larger than that of the
phason, and remains finite in the IC case.

In the presence of an electric field across the stripes, \textit{phason} or
\textit{meandering} fluctuations can be induced. From measurements of the
static dielectric function or conductivity, it should be possible to extract
the amplitude of the charge order parameter. These and other measurements would help constrain parameters in the Ginzburg-Landau model considered here.
Also, direct real-space calculations~\cite{yone} can decrease the number of
independent parameters.

A similar model can be applied to describe commensurate-incommensurate
transitions and response functions in other related classes of materials,
including nickelates and manganites.

This work was supported by the U.S. DOE.

\appendix
\section{single-band Peierls-Hubbard hamiltonian}
\label{app:single}
Here, we consider a single-band Peierls-Hubbard Hamiltonian and estimate the electron-lattice coupling strengths.
The Hamiltonian reads
\begin{eqnarray}
\mathcal{H}&=&-\dsum{ij,\sigma}{}
\Big[t_{ij}+4\gamma'_{ij}\,(w_i+w_j)\Big]c^\dagger_{i\sigma}c_{j\sigma}
\nonumber\\
&&{}+U\dsum{i}{}n_{i\up}n_{i\dn}
+4\gamma\dsum{i,\sigma}{}w_ic^\dagger_{i\sigma}c_{i\sigma}
\nonumber\\
&&{}+\dfrac{1}{2M}\dsum{i}{}(p_{xi}^2+p_{yi}^2)
+\dfrac{K}{2}\dsum{i}{}(u_i^2+v_i^2),
\end{eqnarray}
where
\begin{eqnarray}
w_i&=&\dfrac{1}{4}(u_{i_x,i_y}-u_{i_x-1,i_y}+v_{i_x,i_y}-v_{i_x,i_y-1}),
\end{eqnarray}
$c^\dagger$ ($c$) is the hole creation (annihilation) operator, $t_{ij}$ is the Cu-Cu hopping between the nearest neighbors ($t$) or the next-nearest neighbors ($t'$), and $u_i$ and $v_i$ are the displacements of the $x$-Oxygen and $y$-Oxygen at site $i$, respectively.

To estimate the electron-lattice coupling strength in the single-band model, we start from the three-band (Cu-O) model.
The Cu on-site energy and the Cu-Cu hopping integral to the zeroth order in $t_{pd}$ are given by
\begin{eqnarray}
\epsilon_i^{(0)}=-\Delta_{pd}=-\Delta_0+4\beta w_i, \hspace{.5cm}
\epsilon_{ij}^{(0)}=0.
\end{eqnarray}
The second-order perturbation to the Cu on-site energy and the Cu-Cu hopping
integral are given by
\begin{eqnarray}
\epsilon_i^{(2)}&=&\sum_{\lambda=1,2,3,4}
\frac{t_{pd}(i,\lambda)^2}{\Delta_{pd}(i)},\\
\epsilon_{ij}^{(2)}&=&\frac{1}{2}
\(\frac{t_{pd}(i,\lambda_{ij})t_{pd}(j,\lambda_{ji})}{\Delta_{pd}(i)}
+\frac{t_{pd}(j,\lambda_{ji})t_{pd}(i,\lambda_{ij})}{\Delta_{pd}(j)}\),
\nonumber\\
\end{eqnarray}
where the Cu-O hopping and $d$-$p$ energy difference depend linearly on lattice displacements,
\begin{eqnarray}
t_{pd}(i,\lambda)=t_{pd}\mp \alpha u(i,\lambda),
\hspace{.5cm}
\Delta_{pd}=\Delta_0-4\beta w_i,
\end{eqnarray}
and $\lambda=1,2,3,4$ refer to the oxygen ion to the right of, above, to the left of, and below the Cu ion, respectively; and $\lambda_{ij}$ to that between the $i$ and $j$ Cu ions. By retaining up to linear terms of $u$ and $v$,
$\epsilon_i^{(2)}$ and $\epsilon_{ij}^{(2)}$ are approximately
\begin{eqnarray}
\epsilon_i^{(2)}
&\approx&\frac{4t_{pd}^2 - 8t_{pd}\alpha w_i}{\Delta_0-4\beta w_i}\nonumber\\
&\approx&\frac{4t_{pd}^2}{\Delta_0}
+\(\frac{16 \beta t_{pd}^2}{\Delta_0^2} - \frac{8\alpha t_{pd}}{\Delta_0}\) w_i\\
\epsilon_{ij}^{(2)} &\approx&\frac{1}{2} \(\frac{t_{pd}^2}{\Delta_0-4\beta w_i}
+\frac{t_{pd}^2}{\Delta_0-4\beta w_j}\)\nonumber\\
&\approx& \frac{t_{pd}^2}{\Delta_0} +\frac{2\beta
t_{pd}^2}{\Delta_0^2}(w_i+w_j).
\end{eqnarray}

Combining now the zeroth and second-order results, for the effective one-band model, we obtain
\begin{eqnarray}
\epsilon_i &\approx&-\Delta_0 - \frac{4t_{pd}^2}{\Delta_0}
+ \(4\beta -\frac{16 \beta t_{pd}^2}{\Delta_0^2} +\frac{8\alpha t_{pd}}{\Delta_0}\) w_i\nonumber\\
&\equiv& 4\gamma w_i + \mathrm{const.},\\
\epsilon_{ij}&\approx& -\frac{t_{pd}^2}{\Delta_0} -\frac{2\beta
t_{pd}^2}{\Delta_0^2}(w_i+w_j)
\nonumber\\
&\equiv& -t - 4\gamma'_{ij}(w_i+w_j),
\end{eqnarray}
with the reduced electron-lattice coupling strengths $\gamma$ and $\gamma'$:
\begin{eqnarray}
\gamma&=& \frac{2\alpha t_{pd}}{\Delta_0} + \beta
- \frac{4 \beta t_{pd}^2}{\Delta_0^2},\\
\gamma'_{ij}&=&\frac{\beta t_{pd}^2}{2\Delta_0^2}.
\end{eqnarray}

Using the three-band parameters $t_{pd}=1.3$ [eV], $\Delta=3.6$ [eV],
$0<\alpha<4.5$ [eV/{\AA}], $0<\beta<1$ [eV/{\AA}] and $K=40$
[eV/{\AA$^2$}]~\cite{hybertsen1,yone}, we obtain the single-band parameters:
$t=0.47$ [eV], $0<\gamma<3.73$ [eV/{\AA}] and $0<\gamma'<0.07$ [eV/{\AA}]. Here
the value of the Cu-Cu hopping energy is similar to that estimated in
Ref.~\cite{hybertsen2} ($t=0.43$ [eV]); therefore, it is also expected that
$t'\approx-0.07$ [eV] from Ref.~\cite{hybertsen2}.

\section{estimation of the Ginzburg-Landau parameters}
\label{app:GL-estimate} Here we represent the Ginzburg-Landau parameters for kinetic and pinning energies, $t_0$ and $p_1$, by using the order parameters and other parameters of a single-band Peierls-Hubbard model, and estimate the
commensurability energy and the effective mass of the CDW.

We start with the single-band Hamiltonian in Appendix \ref{app:single}. We can
adopt a particular set of parameters obtained there: e.g., $t=0.47$ [eV],
$t'=-0.07$ [eV], $\gamma=2.0$ [eV/{\AA}], $\gamma'=0$ [eV/{\AA}] and $K=40$
[eV/{\AA$^2$}], etc. However, we will not necessarily require all of these
values below, because it is not our purpose to calculate the order parameters
here. Moreover, the order parameters are assumed to be $\rho \lesssim 10^{-2}$
and $m \lesssim 10^{-1}$ in this paper (\textit{assumptions 4} and \textit{5}).
Our purpose here is to estimate the commensurability energy, the kinetic
energy, and the effective mass of the CDW. In the estimation of the
commensurability energy, we will find that the commensurability energy directly
depends on only $t$, $U$ and $m$ as long as the order parameters are small
($\rho \ll m < t/U$). In the estimation of the kinetic energy and the effective
mass, we will find that those values depend on $\gamma$, $K$ and $\rho$ as long
as $(\rho \frac{\gamma}{K})^2$ is small ($<10^{-5}$ [\AA]). Therefore, we take the parameters as $t=0.47$ eV, $U \ll 20t$, and $\frac{\gamma}{K}<0.1$ [\AA]; and we assume that the calculation with the resulting parameters  yields order parameters which satisfy \textit{assumptions 4} and \textit{5}.
We also consider the terms of $\gamma'$ and $t'$ to be small enough to be neglected,
compared to the $t$ term.

Defining the fluctuations:
\begin{eqnarray}
\begin{array}{ll}
\tilde{n} \equiv n - \<n\>,&
\tilde{u} \equiv u-\<u\>,\\
\tilde{v} \equiv v-\<v\>,&
\tilde{w} \equiv w-\<w\>,
\end{array}
\end{eqnarray}
we rewrite the Hamiltonian:
\begin{eqnarray}
\mathcal{H}&=& \mathcal{H}_{MF}
 + \mathcal{H}_z + \mathcal{H}_{int} + \mathcal{H}_{ph} + \mathcal{H}_{c},
\end{eqnarray}
where
\begin{eqnarray}
\mathcal{H}_{MF}&=&
\dsum{ij,\sigma}{}c^\dagger_{i\sigma}c_{j\sigma}
\Big[ -t_{ij}+ \Big( U\<n_{i\bar{\sigma}}\>
+4\gamma\<w_i\>\Big)\delta_{i,j}\Big],\nonumber\\
\\
\mathcal{H}_{z}&=&
\dsum{i}{}\Big[
4\gamma\dsum{\sigma}{}\tilde{w}_i
\<c^\dagger_{i \sigma}c_{i \sigma}\>
+K\Big(\<u_i\>\tilde{u}_i+\<v_i\>\tilde{v}_i\Big)
\Big],\nonumber\\
\\
\mathcal{H}_{int}&=&
\dsum{i}{}\Big[\,
\dfrac{U}{2}\dsum{\sigma}{}
\tilde{n}_{i\bar{\sigma}}
\tilde{n}_{i\sigma}
+4\gamma\dsum{\sigma}{} \tilde{w}_i \tilde{n}_{i\sigma}
\,\Big],\\
\mathcal{H}_{ph}&=&
\dsum{i}{}\Big[\,
\dfrac{1}{2M}\Big(p_{xi}^2+p_{yi}^2\Big)
+\dfrac{K}{2} \Big(\tilde{u}_i^2+\tilde{v}_i^2\Big)\,\Big],\\
\mathcal{H}_{c}&=&
-U\dsum{i}{}\<n_{i\up}\>\<n_{i\dn}\>
+\dfrac{K}{2}\dsum{i}{}\Big(\<u_i\>^2+\<v_i\>^2\Big).\nonumber\\
\end{eqnarray}
According to the mean-field approximation, we neglect the $\mathcal{H}_{int}$ and $\mathcal{H}_{ph}$ terms.
To consider the equilibrium states, the terms linear in $\tilde{u}_i$ and $\tilde{v}_i$ are set to zero, $\mathcal{H}_{z}=0$.

We assume the order parameters to be described as in Eqs.~(\ref{eq:order-ch}) and (\ref{eq:order-spn}):
\begin{eqnarray}
\<n_{i\up}\> + \<n_{i\dn}\> &=& \rho_0
+\[ \psi(x_i,\phi) + \psi^*(x_i,\phi) \],\\
\<n_{i\up}\> - \<n_{i\dn}\> &=& \[\chi(x_i,\phi) + \chi^*(x_i,\phi)\](-1)^{-\frac{x_i}{a}}\nonumber\\
&=& \chi(x_i,\phi)
\e^{\i \frac{G}{2}x} + \[\chi(x_i,\phi)e^{\i \frac{G}{2}x}\]^*.
\end{eqnarray}
Hence, $\<n_{i\sigma}\>$ can be expanded as
\begin{eqnarray}
\<n_{i\sigma}\>&=&\frac{\rho_0}{2}
+ \<n_{Q\sigma}\>\e^{\i \( Q + \frac{G}{2} \)x_i}
+ \<n_{-Q \sigma}\>\e^{-\i \( Q + \frac{G}{2} \)x_i}\nonumber\\
&&{}+\<n_{2Q\sigma}\>\e^{\i 2Q x_i}+\<n_{-2Q \sigma}\>\e^{-\i 2Qx_i},\\
\<n_{Q\up}\> &=& \<n_{-Q\up}\>^*
={}-\<n_{-Q\dn}\>
=\frac{1}{2\i} m \e^{\i \frac{\phi}{2L}}\\
\<n_{2Q\sigma}\> &=& \<n_{-2Q\sigma}\>^*
=\frac{1}{2} \rho \e^{\i \frac{\phi}{L}},
\end{eqnarray}
where $Q=\frac{q}{2}$.

From $\mathcal{H}_z=0$,
\begin{eqnarray}
\<u_i\>&=&\sum_{\sigma} \frac{\gamma}{K}
\[\<n_{i+1_x\sigma}\>-\<n_{i\sigma}\>\]\\
&=&\frac{\gamma}{K}
\[\psi(x_{i+1_x},\phi)-\psi(x_{i},\phi)\]+\mbox{c.c.}\\
&=& \i \frac{2\gamma}{K} \psi(x_i,\phi) \e^{\i Qa} \sin(Qa)
+\mbox{c.c.}
\end{eqnarray}

We substitute the $Q$-represented form of $\<n_{i\sigma}\>$ and $\<u_i\>$ into $\mathcal{H}_{MF}$, and expand $c$ and $c^\dagger$ as
\begin{eqnarray}
c_{i\sigma}=\frac{1}{\sqrt{N_k}}\sum_{l} \sum_{k_0} c_{k_0+lQ\sigma}
\,\e^{\i (k_0+lQ) r_i},
\end{eqnarray}
where $l$ runs from 0 to $2L-1$ for a commensurability $\frac{1}{L}$, the sum with respect to $k_0$ is taken within the reduced Brillouin zone spanned by $Q$, and $N_k$ denotes the number of $k$ points in the original Brillouin zone.
Hence, $\mathcal{H}_{MF}$ can be rewritten as
\begin{eqnarray}
\mathcal{H}_{MF}&=&
\dsum{k_0,l,\sigma}{}
\Big[ \epsilon(k_0+lQ) c^\dagger_{k_0+lQ \, \sigma}c^{}_{k_0+lQ \, \sigma}\nonumber\\
&&\hspace{.5cm}+\frac{m}{2}U
\(\i\e^{\i \frac{\phi}{2L}}
\,c^\dagger_{k_0+(l+1)Q+\frac{G}{2} \,\sigma}c^{}_{k_0+lQ \,\sigma}
+ \mbox{h.c.}\)\nonumber\\
&&\hspace{.5cm}+\frac{\rho}{2}\tilde{U}
\(\e^{\i \frac{\phi}{L}}
\,c^\dagger_{k_0+(l+2)Q\,\sigma}c^{}_{k_0+lQ\,\sigma} + \mbox{h.c.}\)\Big],\nonumber\\
\end{eqnarray}
where
\begin{eqnarray}
\epsilon(k) &=& -2t\[\cos(k_x a)+\cos(k_y a)\]\nonumber\\
&&-4t'\cos(k_x a) \cos(k_y a) +U\frac{\rho_0}{2}\\
\tilde{U}&=&\[U-\frac{8 \gamma^2}{K}\sin^2\(Qa\)\].
\end{eqnarray}

Only the states near $(\pm\frac{\pi}{2},\pm\frac{\pi}{2})$ at the Fermi level contribute to the condensation energy.  The energy gap around
$(\pm\frac{\pi}{2},\pm\frac{\pi}{2})$ is approximately given by
\begin{eqnarray}
E_{g} \approx \sqrt{2\Delta^2 + \frac{2\Delta^{2L}}{D^{2L-2}}\cos(\phi)},
\end{eqnarray}
where $\Delta\equiv\frac{mU}{2}$ and $D \sim 2t$.
The approximation here is justified when $D\gg\Delta$ and $mU \gg \rho\tilde{U}$.
This is the case for $L=4$ and $U \ll 20t$ under \textit{assumptions 4} and \textit{5}.
Therefore, the commensurability effect on the condensation
energy, $\delta_C$, is given by
\begin{eqnarray}
\delta_{\C}^2 &=& \frac{2\Delta^{2L}}{D^{2L-2}}.
\end{eqnarray}

Now we consider the energy to be calculated from the meanfield Hamiltonian
\begin{eqnarray}
\mathcal{H}=\mathcal{H}_{MF}+\mathcal{H}_{c}.
\end{eqnarray}
In the energy, the component proportional to $m^2$ contributes to the formation of the ordered state as well as the $r_0$ term in the Ginzburg-Landau free energy (\ref{eq:f0}).
The condensation energy per site is given by
\begin{eqnarray}
E_{con}(\phi) =  \frac{2\Delta^2}{U} - \frac{E_g^2}{\Xi},
\end{eqnarray}
where the first term is from $\mathcal{H}_{c}$, and the second from $\mathcal{H}_{MF}$;
and $\Xi$ depends on the model system parameters (bandwidth, interactions, etc.) and doping.
Note that $E_{con}(\phi)$ should have two components which are, respectively, proportional to $m^2$ and $m^{2L}$.
The former is the $\phi$-independent condensation energy, $-E_{con}^{(0)}$, and the latter is the commensurability energy, $-E_{\C}$.
Setting $\phi=\frac{\pi}{2}$, $E_{con}^{(0)}$ can be written by
\begin{eqnarray}
E_{con}^{(0)} = 2\Delta^2 \(\frac{1}{\Xi} - \frac{1}{U}\)
=\frac{1}{2}|r_0|m^2.
\end{eqnarray}
The commensurability energy is given by
\begin{eqnarray}
E_\C &=& E_{con}(\mbox{$\frac{\pi}{2}$}) - E_{con}(0)
= \Xi^{-1} \delta_c^2\\
&=& \(U+|r_0| \)\(\frac{U}{D}\)^{2L-2}\frac{m^{2L}}{2^{2L-1}}
\end{eqnarray}

For an insulating CDW in the nearly-half-filled one-dimensional system, the condensation energy is roughly estimated as $\sim E_g^2/t$, so that we expect
\begin{eqnarray}
|r_0| \lesssim \frac{U^2}{t} \sim \mbox{$U^2$ [eV]}
\end{eqnarray}
for the metallic CDW in a two-dimensional system. Since we consider $U\ll 10$
[eV], $U+|r_0|$ should be on the order of $U$ [eV]. In this case, the
commensurability energy per site is obtained by
\begin{eqnarray}
E_{\C} &\sim& \mbox{$U^7 m^8 \times 10^{-2} $ [eV]}.
\end{eqnarray}
Then, we obtain the Ginzburg-Landau parameter for pinning energy
\begin{eqnarray}
|\tilde{p}_1| &\sim& \mbox{$U^7 \Lambda^{-4} \times 10^{-2} $ [eV]},
\end{eqnarray}
where $\Lambda=\frac{\rho}{m^2}$.

Next, to estimate the kinetic energy, we consider the time-dependence of the order parameter through $\phi$
\begin{eqnarray}
\phi(x,t)&=& \phi_0 + \xi(t)\cos(kx+\phi_1).
\end{eqnarray}
The displacements are written as
\begin{eqnarray}
u_i(t)&=&\mathcal{U}_i(t)+\mathcal{U}_i^*(t),\\
\mathcal{U}_i(t)&=&\i\frac{2\gamma}{K} \psi(x_i,\phi) \e^{\i qa/2}\sin\(\frac{qa}{2}\).
\end{eqnarray}

\begin{eqnarray}
&&\left|\frac{d \,\mathcal{U}_i}{dt}\right|^2
=\(\frac{2\gamma}{K}\)^2 \sin^2\(\frac{qa}{2}\)
\left|\frac{d\psi}{dt}\right|^2\\
&&=\(\frac{2\gamma}{K}\)^2 \sin^2\(\frac{qa}{2}\)
q^2\rho^2v_{d}^2\cos^2(kx_i+\phi_1),
\end{eqnarray}
where $v_{d}=\frac{d\xi}{q L dt}$.
The mean square velocity of the ion and the total kinetic energy are
\begin{widetext}
\begin{eqnarray}
\overline{\left|\frac{d \,\mathcal{U}_i}{dt}\right|^2} =\frac{1}{N}\sum_{i}
\left|\frac{d \,\mathcal{U}_i}{dt}\right|^2 =\(\frac{2\gamma}{K}\)^2
\sin^2\(\frac{qa}{2}\) q^2\rho^2v_{d}^2\frac{1+\cos(2\phi_1)\delta_{k,0}}{2},
\end{eqnarray}
\begin{eqnarray}
\frac{1}{2}\rho_0 \mu v_{d}^2 &\equiv&
\frac{1}{2}\rho_0m_ev_{d}^2+\frac{1}{2}M
\overline{\left|\frac{d \,\mathcal{U}_i}{dt}\right|^2}
=\frac{1}{2}\rho_0m_ev_{d}^2
\[1+\frac{M}{\rho_0 m_e}\(\frac{2\gamma}{K}\)^2 \sin^2\(\frac{qa}{2}\)
q^2\rho^2\frac{1+\cos(2\phi_1)\delta_{k,0}}{2}\],
\end{eqnarray}
\end{widetext}
where $M$, $m_e$, and $\mu$ are the ionic mass per site, the electron mass, and the effective mass of a hole in the CDW, respectively.

On the other hand, the kinetic energy in the Ginzburg-Landau expansion is written by
\begin{eqnarray}
T &=& t_0 \int \frac{dx}{V}\left|\frac{d\psi}{dt}\right|^2\\
&=& t_0 q^2\rho^2 \frac{1+\delta_{k,0}\cos(2\phi_1)}{2} v_d^2\\
&\equiv& \[\tau_0 + \tau_1 q^2\rho^2 \frac{1+\delta_{k,0}\cos(2\phi_1)}{2}\] v_d^2,
\end{eqnarray}
where both $\tau_0$ and $\tau_1$ are independent of $\rho$.
Near $\frac{1}{8}$ doping, $\tau_0$ and $\tau_1$ are estimated as
\begin{eqnarray}
\tau_0 &=& \frac{1}{2} \rho_0 m_e \approx 2.9\rho_0\times10^{-32} \,\,\mbox{[eV s$^2$/{\AA}$^2$]}\\
\tau_1 &=& \frac{M}{2}\(\frac{2\gamma}{K}\)^2 \sin^2\(\frac{qa}{2}\)\\
&\approx& 1.7\(\frac{\gamma}{K}\)^2 \times 10^{-27}\,\,\mbox{[eV s$^2$]}
\end{eqnarray}
Therefore, the Ginzburg-Landau parameter for the kinetic energy, $t_0$, is given by (for $k=0$ and $\phi_1=0$)
\begin{eqnarray}
t_0 &\approx& \[\frac{33.7 \rho_0 + 1.7\rho^2 \(\frac{\gamma}{K}\)^2 \times10^5 }{\rho^2} \] \times 10^{-32}\,\,\mbox{[eV s$^2$]}\nonumber\\
&\approx& 4.2 \rho^{-2}\times10^{-32}.
\end{eqnarray}
where we use $\frac{\gamma}{K}<10^{-1}$ and the \textit{assumptions 4} and \textit{5}.
The effective mass is
\begin{eqnarray}
\mu &\approx& 2\tau_0/\rho_0 + \tau_1 q^2\rho^2 \[1+\delta_{k,0}\cos(2\phi_1)\]/\rho_0
\end{eqnarray}
In the case of $k=0$ and $\phi_1=0$,
\begin{eqnarray}
\mu&\approx& \[\frac{9.2\rho_0 + 4.8\rho^2\(\frac{\gamma}{K}\)^2\times10^{4}}{\rho_0}\] \times 10^{-31}\,\,\mbox{[kg]}\nonumber\\
&\approx& 9.2 \times 10^{-31}\,\,\mbox{[kg]}.
\end{eqnarray}
For $\rho < 10^{-1}$, the electronic energy is dominant in the total kinetic energy.

\section{Equation of motion for CDW in Electric field}
\label{app:eq-motion} Here, we derive the equation of motion for a meandering CDW oscillating in the $x$-direction in an ac electric field (discussed in Sec.~\ref{sec:electric-field}).
We consider a plane electromagnetic wave with the electric field oriented in the $x$-$y$ plane, $\mathbf{E}=E_0 (k_y/k,-k_x/k)\, e^{\i(k_x x + k_y y) -\i\omega t}$.
The corresponding vector potential is $\mathbf{A} = \mathbf{E}/(-\i\omega)$.

The vector potential couples to the electrical current $\mathbf{j}$, which can be
expressed in terms of the carrier density and the velocity.  The velocity of
the density wave can be expressed via the time derivative of this phase, $v_x =
(1/G)\dot{\varphi}$. Thus the form of the phase that couples to the plane-wave
electric field as above is
\begin{eqnarray}
\varphi(x,y,t) &=& \xi(y,t) \, e^{\i k_x x }
\end{eqnarray}

Here we take $k_x \rightarrow 0$.
Thus, the additional contribution to the Lagrangian due to the interaction with
the external electromagnetic field is $\mathbf{A}\cdot\mathbf{j} = \frac{e\rho_0}{a^2cG} A_x\dot{\xi}$.
Combining it with the bare Lagrangian, Eq.~(\ref{eq:L-meandering}), we obtain
\begin{eqnarray}
\mathcal{L} &=&  \frac{t_0\rho^2}{L^2}\int \frac{dy}{V_y}
 \Bigg[
 \(\frac{d\xi}{dt}\)^2
-\frac{\tilde{h}_0}{t_0}\(\frac{d \xi}{dy}\)^2
-\Delta_\phi^2\xi^2\Bigg] \nonumber\\
&& - \int\frac{dy}{V_y} \frac{e\rho_0 }{G} \frac{E_x}{(-\i\omega)}\dot{\xi}.
\end{eqnarray}
Adding a dissipation term, we obtain the equation of motion:
\begin{eqnarray}
\frac{d^2\xi(y,t)}{dt^2}+\Gamma\frac{d\xi(y,t)}{dt}
-\frac{\tilde{h}_0}{t_0}\frac{d^2 \xi}{dy^2}
+\Delta_\phi^2\xi(y,t)\nonumber\\
=\frac{eG}{\mu}E_x(y,t).
\end{eqnarray}

\end{document}